\documentclass[12pt,preprint]{aastex}

\newcommand{\Difd}[2]{{d #1 \over d #2}}
\newcommand{\Dif}[2]{{d #1 / d #2}}
\newcommand{\be}{\begin{equation}}
\newcommand{\ee}{\end{equation}}

\shorttitle{Structure of Filamentary Clouds}
\shortauthors{Tomisaka}

\begin{document}
\title{Polarization Structure of Filamentary Clouds}
%\title{Polarization Structure of Filamentary Cloud}
\author{Kohji Tomisaka\altaffilmark{1}}
\affil{Division of Theoretical Astronomy, National Astronomical Observatory of Japan,
Mitaka, Tokyo 181-8588, Japan}
\email{tomisaka@th.nao.ac.jp}
\altaffiltext{1}{also at Department of Astronomical Science, 
School of Physical Sciences,
SOKENDAI (The Graduate University for Advanced Studies),
Mitaka, Tokyo 181-8588, Japan}
\begin{abstract}
%Filaments are considered to be basic structures that consist of interstellar molecular clouds.
Filaments are considered to be basic structures and molecular clouds consist of filaments.
Filaments are often observed as extending in the direction perpendicular to the
 interstellar magnetic field.
The structure of filaments has been studied based on a magnetohydrostatic equilibrium
 model \citep{tomisaka2014}.   
Here, we simulate the expected polarization pattern for isothermal magnetohydrostatic filaments.
The filament exhibits a polarization pattern in which the magnetic field is
 apparently perpendicular to the filament when observed from the direction perpendicular 
 to the magnetic field.
When the line-of-sight is parallel to the global magnetic field,
 the observed polarization pattern is dependent on the center-to-surface density ratio for the filament and the concentration of the gas mass toward the central magnetic flux tube.
Filaments with low center-to-surface density ratios have an
 insignificant degree of polarization when observed from the direction parallel to the global magnetic field.
However, models with a large center-to-surface density ratio
 have polarization patterns that indicate the filament is perpendicularly threaded by the magnetic field.
When mass is heavily concentrated at the central magnetic flux tube,
 which can be realized by the ambipolar diffusion process,
 the polarization pattern is similar to that expected
 for a low center-to-surface density contrast.
\end{abstract}

\section{Introduction}
Filamentary clouds have been attracting much attention
 since the {\it Herschel} satellite identified many filaments
 in interstellar molecular clouds \citep{menshchikov2010,miville2010,arzoumanian2011,hill2011,
 schneider2012}.
Some filamentary clouds are composed of multiple sub-filaments which are also coherent in velocity
 space \citep{hacar2013}.
These filaments are beginning to be considered as one of the building blocks
 of interstellar gas and thus must have an important 
 role in the star formation process.\footnote{
In addition, filamentary objects are also found in our diffuse interstellar medium; 
 in the ionized medium, \citet{gaensler2011} and \citet{iacobelli2014} found filamentary structure
 in the map of polarization gradients. 
 In the diffuse Galactic H{\sc I}, slender, linear features are found \citep{clark2014},
 which extend in the direction of interstellar magnetic field.
Thus, we encounter filamentary structures in various phases of interstellar gas.}
The relationship between magnetic fields and filaments has been studied by observations
 of interstellar near IR polarization \citep{sugitani2011,palmeirim2013},
 and was explained by dichroic extinction due to dust grains aligned with the magnetic field.
These observations indicate that the filaments are extending in the direction perpendicular
 to the interstellar magnetic field.

Polarization observations with {\it Planck} at 353 GHz give us more statistical view on the relationship
 between the magnetic field and the structure of the molecular clouds \citep{planck2015}.
The angle ($\varphi$) has been calculated pixel by pixel
 between the projected interstellar magnetic field and the direction of iso-column density contours.
In typical molecular clouds such as Taurus, Lupus, and Chamaeleon-Musca,
 distribution of this angle peaks around $\varphi\sim \pm 90^\circ$
 for high-density regions with the column density larger than $N_{\rm H}\gtrsim 10^{22} {\rm cm^{-2}}$.
This means that
 magnetic field is observed preferentially perpendicular to the filament
 with $N_{\rm H}\gtrsim 10^{22} {\rm cm^{-2}}$.\footnote{ 
Similar analysis is also tried for more small-scale structures with use of SMA \citep{koch2013}.}
Relation between the intercloud magnetic field and major axes of the filamentary clouds 
 is studied for Gould Belt clouds by \citet{li2013}.
Although another sequence of clouds is proposed, in which 
 the directions of the filament extension
 and the intercloud magnetic fields are parallel,
 the same perpendicular configuration is also confirmed\citep{li2013}. 

As for the Serpens South Cloud, the magnetic field seems running perpendicular to the
 long axis of the molecular cloud \citep{sugitani2011}, that is $\varphi\sim 90^\circ$.
Using multi-line observations, \citet{kirk2013} estimated the accretion rate onto an embedded 
 cluster-forming region in this cloud: $\sim 30\,M_\odot\,{\rm yr^{-1}}$ is accreting along the axis of the filament while
 $\sim 130\, M_\odot\,{\rm yr^{-1}}$ is radially contracting (see also Figure 9 of \citet{andre2014}).
This large accretion rate is consistent with the fact that the observed mass per unit length (line-mass)
 $\lambda \sim 60\, M_\odot\,{\rm pc^{-1}}$  \citep{kirk2013} exceeds the critical
 line-mass of thermally supported filaments at 10K, $\lambda_{\rm th\ crit}\simeq 16.7\,M_\odot\,{\rm pc^{-1}}$ \citep{stodolkiewicz1963,ostriker1964,inutsuka1997}.
%Although the filaments may be in a dynamical contraction stage \citep{inoue2013},
% there is a possibility that the observed filaments are in hydrostatic equilibrium.
%We have previously studied isothermal filaments that are threaded by the interstellar 
% magnetic field under magnetohydrostatic conditions (\citet{tomisaka2014}; hereafter paper I).
The magnetically critical line-mass of the isothermal filaments that are perpendicularly 
 threaded by the interstellar magnetic field is studied  by \citeauthor{tomisaka2014}
 (\citeyear{tomisaka2014}; hereafter paper I) under magnetohydrostatic conditions.
This shows that the magnetic field can support the filament against the self-gravity,
 as long as the line-mass %(mass per unit length)
 of the filament is
 $\lesssim 0.24 \Phi_{\rm cl}/G^{1/2}$, where $\Phi_{\rm cl}$ and $G$ represent
 one half of the magnetic flux threading the filament per unit length and the gravitational constant,
 while the line mass is limited below $\lambda_{\rm th\ crit}=2\,c_s^2/G$ for a filament with no magnetic field
 ($c_s$ represents the isothermal sound speed).
Since the magnetically critical line-mass is given $\lambda_{\rm mag\ crit}\simeq 22.4\,M_\odot\,{\rm pc^{-1}}
 (R_0/0.5{\rm pc})(B_0/10\mu{\rm G})$ (equation (39) of paper I), %\citet{tomisaka2014}),
 where the radius of the filament $R_0$ and the field strength $B_0$ give the amount of magnetic
 flux threading the filament, the filament of the Serpens South Cloud may
 be magnetically supercritical, $\lambda > \lambda_{\rm mag\ crit}$ not only thermally supercritical,
 $\lambda > \lambda_{\rm th\ crit}$.
That is, when the magnetic flux per unit length is sufficiently large, such as
 $\Phi_{\rm cl} \gtrsim 3\,{\rm pc\,\mu G}\,(c_s/190\,{\rm m\,s^{-1}})^2$,
 the magnetic field plays a crucial role to support the filament.

The relationship between magnetic field and the direction of the major axis of filaments
 is believed to be related to formation mechanisms.
There are several mechanisms to form filamentary clouds.   
\citet{nagai1998} considered an isothermal sheet with uniform magnetic fields
 and its fragmentation to filaments.
They obtained filaments perpendicular to the magnetic field. 
When clouds contract along the magnetic field lines by the self-gravity,
 major axes of the filaments tend to align in the perpendicular direction to the magnetic field
 \citep[e.g.][]{nakamura2008}.
Several models of MHD turbulence are proposed to form filamentary structures \citep{padoan2014}.    
Sub-Alfv\'{e}nic anisotropic turbulence leads to filaments to be aligned along the
 magnetic field lines \citep{stone1998}.
On the other hand, in super-Alfv\'{e}nic turbulence, shocks form thin sheets \citep{padoan2001}.
Magnetic fields are also compressed in the sheets and as a result
 the field direction is parallel to the filamentary feature which comes from the compressed sheet. 
\citet{inoue2013} considered collisions between two magnetized molecular clouds. 
The deformed MHD shock wave kinks the stream lines, and accumulates molecular gas into a filament 
 extending perpendicular to the magnetic field (see their Figure 1).
However, to discuss the formation mechanism, we have to reconstruct three-dimensional configuration
 of magnetic fields and filaments from two-dimensional polarization maps.

The degree of polarization is low if the object is observed from the direction parallel to 
 the magnetic field, 
 either for interstellar polarization due to the dichroic extinction in the
 optical and infrared wavelengths,
 or for the polarization of thermal emissions from dust grains aligned with the magnetic field.
If the magnetic field is threading the filament perpendicularly, then an 
 appreciable number of such objects must be observed as weakly polarized objects.
However, observed examples of filaments indicate that the magnetic field is perpendicular to 
 the filament. 
The polarization is affected by integration along the line-of-sight; therefore,
 it is not so simple to estimate the polarization pattern only from the
 angle between the local magnetic field and the line-of-sight. 
%Thus, the expected polarization for such filaments can be calculated and 
% the structure of the magnetohydrostatic filaments discussed, especially 
% their magnetic structure as expected from the polarization pattern.
Thus, we calculate the expected polarization for such filaments and 
 discuss the structure of the magnetohydrostatic filaments, especially, 
 their magnetic structure expected in the polarization pattern.

The structure of this paper is as follows.
 Models of magnetohydrostatic filaments are taken from paper I.
 The models and formulation to calculate polarization are shown in $\S$2.
In $\S$3, the expected polarization is given for two typical filament models;
 one with a relatively low center-to-surface density ratio, $\rho_c/\rho_s=10$,
 and another with a relatively high ratio, $\rho_c/\rho_s=300$.
These two models exhibit distinctly different polarization patterns.
Section 4 is devoted to discussion and exploration of the structures of filaments
 with different mass loadings (mass distribution against magnetic flux tube).
In $\S$4, expected distributions of $\varphi$ are also calculated 
 for magnetohydrostatic filamentary clouds.

\section{Method}
Here, we focus on the polarization expected in the thermal dust emissions. 
Assuming an infinitely long filament, the magnetohydrostatic structure
 is specified with three nondimensional parameters (paper I): 
 the center-to-surface density ratio, $\rho_c/\rho_s$,
 the plasma beta of the ambient material from far outside
 the cloud, $\beta_0 \equiv \rho_sc_s^2/(B_0^2/8\pi)$,
 and the radius of a `parent' filament normalized with the scale-height,
 $R_0/[c_s/(4\pi G\rho_s)^{1/2}]\equiv R_0'$,
 where the parent filament is a virtual state from which the filament
 is formed under magnetic flux freezing.
We assume no additional turbulence motion in the filament.
In these definitions, $\rho_s$ represents
 the density at the surface of the filament, 
 outside of which a tenuous medium with a pressure of $\rho_s c_s^2$ is distributed,
 where $c_s$ and $G$ indicate the isothermal sound speed and gravitational
 constant, respectively.
Beside these three scalar parameters,
 to specify a solution for magnetohydrostatic equilibrium,
 the distribution of magnetic flux against mass, which has a freedom of function,
 must be constrained. 
In paper I, we assumed a magnetic flux distribution which is realized when
 a uniform-density cylinder with a radius $R_0$ is threaded with a uniform magnetic field,
 $B_0$.
The Cartesian coordinate system is used, where 
 the filament is extending in the $z$-direction and the global magnetic field is running in the $y$-direction
 (see Figure 1 of paper I).
The density distribution $\rho$, and magnetic field lines for equilibrium structures in the $x$-$y$ plane
 are shown in Figures 2, 5, and 7 of paper I. 
The density $\rho (x,y)$, and magnetic field ${\bf B}(x,y)=(B_x(x,y), B_y(x,y))$,
 are uniform in the $z$-direction and are dependent only on $(x,y)$. 

\begin{figure}
\epsscale{.50}
%\plotone{ps/coordinate.eps}
\plotone{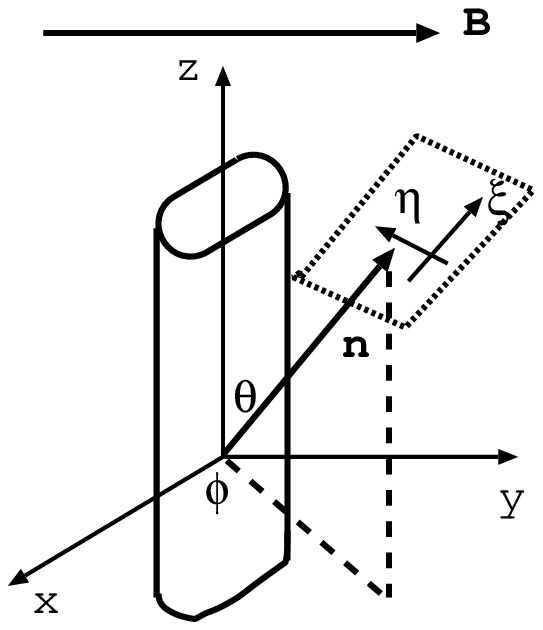}
\epsscale{.80}
\caption{\label{fig:coordinate}
Schematic view of the geometry.
Magnetohydrostatic filament extending in the $z$-direction is
 threaded by a magnetic field that runs globally in the $y$-direction.
The filament is symmetric with respect to both the $x=0$ and $y=0$ planes.
The observation line-of-sight is directed along a unit vector {\bf n},
 of which the direction is specified by two angles, $\theta$ and $\phi$.
The observation is drawn on another plane with the coordinate $(\xi,\eta)$,
 of which the direction is defined in Equations~(\ref{eqn:eta}) and (\ref{eqn:xi}).
The filament is uniform in $z$-direction; therefore,
 the result is dependent only on $\xi$.}
\end{figure}

When an object is observed along a line-of-sight whose direction is specified by a unit vector
 ${\bf n}$,
 another Cartesian coordinate is introduced to indicate the observation, $(\xi, \eta)$, of which
 the unit vectors are as follows:
\begin{mathletters}
\begin{eqnarray}
 {\bf e}_\eta&=&\frac{{\bf e}_z-({\bf e}_z\cdot {\bf n}){\bf n}}
 {|{\bf e}_z-({\bf e}_z\cdot {\bf n}){\bf n}|},\label{eqn:eta}\\
 {\bf e}_\xi&=&{\bf e}_\eta\times{\bf n},\label{eqn:xi}
\end{eqnarray}
\end{mathletters}
where the above definitions are the same as those given in \citet{tomisaka2011}.
The geometry of the filament and the direction of observation are shown in Figure \ref{fig:coordinate}. 
The polarization of the thermal dust emissions is calculated from the relative Stokes parameters
 \citep{lee1985,fiege2000,matsumoto2006,tomisaka2011,padovani2012}:
\begin{mathletters}
\begin{eqnarray}
 q&=&\int \rho\cos 2\psi \cos^2\gamma ds, \label{eqn:stokes-q}\\
 u&=&\int \rho\sin 2\psi \cos^2\gamma ds, \label{eqn:stokes-u}
\end{eqnarray}
\end{mathletters}
 where the integration is performed along the line-of-sight,
 $\rho$ is the density,
 and $\gamma$ and $\psi$ represent
 the angle between the magnetic field and the celestial plane, and
 the angle between the $\eta$-axis and the magnetic field projected on the celestial plane, respectively
 (see Figure 3 of \citet{tomisaka2011}).
The E-vector distribution obtained from the polarimetry of background stars in the optical/near infrared regions
 appears similar to the polarization B-vector expected for the thermal dust emissions obtained here,
 except when the optical depth is thick.
The dust temperature and the degree of dust alignment with the magnetic field may change
 spatially\footnote{
Mechanisms to align dust grains along magnetic fields have been a long-standing problem
 under debate.
Suprathermal rotation \citep{purcell1979} achieved by e.g. the radiation torque
 \citep{draine1996,hoang2009} maintains its alignment for long time \citep{draine2011}.
Magnetization of rotating uncharged dust grains by Barnett effect (\citeyear{barnett1915}) 
 is believed to be efficient in dust alignment process.
Thus, we assume here the degree of dust alignment is uniform.}. 
However, in the present calculation, we assume that the dust temperature and
 the degree of alignment are spatially uniform.
Observation of the filament from the direction of the magnetic field
 yields $\gamma=90^\circ$ (the magnetic field is perpendicular to the celestial plane).
This configuration contributes nothing to the relative Stokes parameters; therefore,
 the observed degree of polarization is low when observed from the magnetic direction.
The polarization direction $\chi$, is calculated from the relative Stokes parameters, 
 $q$ and $u$ of Equations (\ref{eqn:stokes-q}) and (\ref{eqn:stokes-u}) as
\begin{mathletters}
\begin{eqnarray}
\cos 2\chi &=&\frac{q}{(q^2+u^2)^{1/2}},\\
\sin 2\chi &=&\frac{u}{(q^2+u^2)^{1/2}}, 
\end{eqnarray}
\end{mathletters}
 which gives the vector for the degree of polarization,
\be
 {\bf P}=\left(\begin{array}{c}P_\xi\\
                               P_\eta
               \end{array} \right)
=\left(\begin{array}{c}P\sin\chi\\
                       P\cos\chi
               \end{array} \right).
\ee
The polarization degree $P$ is calculated relatively empirically:
\be
 P=p_0\frac{(q^2+u^2)^{1/2}}{\Sigma-p_0\Sigma_2},
\ee      
with use of the following integrated quantities: 
\begin{eqnarray}
 \Sigma&=&\int \rho \ ds,\label{eqn:Sigma}\\
 \Sigma_2&=&\int \rho \left(\frac{\cos^2\gamma}{2}-\frac{1}{3}\right)ds.\label{eqn:Sigma2}
\end{eqnarray}
The parameter $p_0$ controls the maximum degree of polarization and
 we assume $p_0=0.15$ to fit the highest degree of polarization 
 observed for the interstellar cloud.

The path length $\Delta s$, crossing one grid cell along the line-of-sight
 (used in Equations (\ref{eqn:stokes-q}), (\ref{eqn:stokes-u}),
 (\ref{eqn:Sigma}), and (\ref{eqn:Sigma2})) is calculated from
 the two-dimensional path length in the $x$-$y$ plane $\Delta \ell$, as
\be
 \Delta s=\Delta \ell / \cos \theta,
\ee
where $\cos \theta$ represents the direction cosine of the line-of-sight to the $z$-direction. 
Equation (\ref{eqn:Sigma}) contains only $\rho(x,y)$ in the integrand,
 so that $\Sigma$ is proportional to $(\cos \theta)^{-1}$.
Other integrands also contain $\gamma$ and $\psi$,
 which are dependent on the line-of-sight direction or $\theta$ and $\phi$,
 where the spherical coordinate $(\theta,\phi)$ is adopted
 to specify the direction of the line-of-sight.
The polarization distributions expected for the respective models are calculated
 with different $\theta$ and $\phi$.

\section{Results}
\label{sec:result}
\begin{table}
\begin{tabular}{lcccc}
Model & $R_{0}$ & $\beta_0$ & $\rho_c$ & $\lambda_0$\\
\hline
A\ldots\ldots & $2\,c_s/(4\pi G \rho_s)^{1/2}$ & 1 & $10\,\rho_s$ & $1.71 \,c_s^2/G$\\
B\ldots\ldots & $2\,c_s/(4\pi G \rho_s)^{1/2}$ & 1 & $300\,\rho_s$& $2.26 \,c_s^2/G$\\
\hline
\end{tabular}
\caption{\label{tbl1:model-parameters}{\small Model parameters of Figure \ref{fig:Paper1Fig5}. 
$R_0$, $\beta_0$, $\rho_c$, and $\lambda_0$ represent
 the radius of a `parent cloud' from which the filament is formed under magnetic
 flux freezing, the plasma beta of the ambient material from far outside
 the cloud, the density at the center of the filament, and 
 the mass per unit length of the filament, respectively.} 
}
\end{table}

We calculate the polarization pattern for filaments in magnetohydrostatic 
 balance obtained in paper I,
 of which the structures are shown in Figure \ref{fig:Paper1Fig5}. 
Figures \ref{fig:Paper1Fig5}(a) and (b) show typical models
 with low density contrast $\rho_c/\rho_s=10$,
 and with high density contrast $\rho_c/\rho_s=300$, respectively.
The respective line-masses of the filaments are equal to
 $\lambda_0=1.71 c_s^2/G=22 c_s^2/(4\pi G)$ and $\lambda_0=2.26c_s^2/G=28c_s^2/(4\pi G)$ 
(model parameters are summarized in Table \ref{tbl1:model-parameters}).
Figure 2 shows that magnetic field is relatively uniform in the solution 
 with low $\rho_c/\rho_s$ (Model A), while the magnetic field lines are strongly squeezed
 near the equator ($y=0$), when $\rho_c/\rho_s$ is high (Model B).
To show the polarization distribution,
 we assume $c_s=0.19 {\rm km\,s^{-1}}$,
 $\rho_s=10^3 {\rm H_2\,cm^{-3}}$,
 and thus the scale-height $c_s/(4\pi G \rho_s)^{1/2}=3.1\times 10^4 {\rm AU}$. 

\subsection{Model with Low Central Density}
Figure \ref{fig:rhoc10pol} shows the polarization pattern expected for
 Model A of Figure \ref{fig:Paper1Fig5}(a) with low density contrast,
 $\rho_c/\rho_s=10$.
In the present paper, we show the direction of the B-vector for the observed electromagnetic wave
 as the direction of polarization, which coincides with the direction of the interstellar magnetic field
 when the temperature, density and magnetic field are all uniform. 
Observation of the filament from near its axis ($\theta=30^\circ$:
 Figures~\ref{fig:rhoc10pol}(a)-(c)) indicates
 that the polarization direction is dependent on the azimuthal angle, $\phi$. 
When observing the filament from $(\theta,\phi)=(30^\circ,0^\circ)$, which is
 a direction in the $x-z$ plane (perpendicular to the global magnetic field),
 the polarization vector is perpendicular to the filament
 (Figure~\ref{fig:rhoc10pol}(a)).
However, when observing from a direction in the $y-z$ plane, 
 such as $(\theta,\phi)=(30^\circ,90^\circ)$, the polarization vector is
 parallel to the filament (Figure~\ref{fig:rhoc10pol}(c)).
Between these two, the polarization vector is directed from the upper-left 
 to the lower-right (Figure~\ref{fig:rhoc10pol}(b)).
The degree of polarization decreases when we increase $\phi$ from $\phi=0^\circ$ to $\phi=90^\circ$.
This is reasonable because observation of the target from the direction of the magnetic field
 induces a low degree of polarization.  

This is clarified by a comparison of three models with $\theta=80^\circ$
 (Figures~\ref{fig:rhoc10pol}(d)-(f)).
The direction $(\theta,\phi)=(80^\circ,0^\circ)$ is almost perpendicular
 to the global magnetic field (Figure~\ref{fig:rhoc10pol}(d)),
 while $(80^\circ,90^\circ)$ is almost parallel to it (Figure~\ref{fig:rhoc10pol}(f)).
Observation from $\theta \simeq 90^\circ$ shows that the polarization vector is 
 perpendicular to the filament, even for $\phi=45^\circ$.
In this configuration, the degree of polarization is extremely low, when 
 the filament is observed from near the magnetic field direction (Figure~\ref{fig:rhoc10pol}(f)).
Thus, in the models shown in Figure \ref{fig:rhoc10pol},
 the polarization pattern coincides with that expected for the models
 consisting of a uniform magnetic field and uniform-density dust distribution.
 
Figure \ref{fig:rhoc10pol_all} shows
 the polarization angle (Figures~\ref{fig:rhoc10pol_all}(a) and (d)),
 the column density (Figures~\ref{fig:rhoc10pol_all}(b) and (e)),
 and the degree of polarization (Figures~\ref{fig:rhoc10pol_all}(c) and (f))
 against the $\xi$-axis, which is taken to be perpendicular to the filament
 (see Figure~\ref{fig:coordinate}). 
The upper and lower panels correspond
 to the cases of $\theta=30^\circ$ and $\theta=80^\circ$, respectively.

In Figures \ref{fig:rhoc10pol_all}(a) and (d), $\alpha$, the angle between the filament axis and 
 the polarization {\bf B} vector are plotted.
$\alpha=90^\circ$ indicates the polarization direction is perpendicular to the filament, 
 while $\alpha=0^\circ$ and $\alpha=180^\circ$ indicate that the polarization direction and the filament are parallel.
In Figure \ref{fig:rhoc10pol_all}(a), the polarization angle increases
 from $\alpha\sim 90^\circ$ at $\phi=0^\circ$ (lower solid line; Figure~\ref{fig:rhoc10pol}(a))
 to   $\alpha\sim 180^\circ$ at $\phi=90^\circ$ (upper solid line; Figure~\ref{fig:rhoc10pol}(c)).
As $\phi$ increases from $0^\circ$ to $90^\circ$,
 a deviation from the direction perpendicular to the filament appears first
 for the line-of-sight passing through the center, $\xi=0$.
Figure \ref{fig:rhoc10pol_all}(d) shows the models with $\theta=80^\circ$.
The polarization angle $\alpha$ increases from $90^\circ$ to $180^\circ$
 when changing the azimuth angle of the line-of-sight $\phi$,
 from $0^\circ$ to $90^\circ$, similar to that in Figure~\ref{fig:rhoc10pol_all}(a). 
Although the polarized intensity is weak for models with $\phi\gtrsim 60^\circ$
 (Figure~\ref{fig:rhoc10pol_all}(f)),
 the polarization vector is within a $\pm 10^\circ$ deviation from the perpendicular direction (Figure~\ref{fig:rhoc10pol_all}(d)).

Figures \ref{fig:rhoc10pol_all}(b) and (e) show the column density distribution for two groups with 
line-of-sights of $\theta=30^\circ$ and $80^\circ$, respectively.
$\Sigma \propto (\cos \theta)^{-1}$; therefore, the column density distribution is
 scaled between two models of $\theta=30^\circ$ and $\theta=80^\circ$.
This filament has a major axis in the $x$-axis (Figure~\ref{fig:Paper1Fig5}),
 so that the width of $\Sigma$ distribution is observed to be narrower for the line-of-sight with
 $\phi=0^\circ$ and wider for $\phi=90^\circ$.

Figures \ref{fig:rhoc10pol_all}(c) and (f) show the expected degree of polarization, $P$, which is dependent on $\theta$;
 when the filament is observed from the direction of the filament axis,
 a larger polarization intensity is expected (Figure~\ref{fig:rhoc10pol_all}(c)).
For the line-of-sight of $\theta=30^\circ$,
 a relatively high degree of polarization, $10\% \lesssim P \lesssim 15\%$, is 
 observed, irrespective of $\phi$.
However, for the line-of-sight of $\theta=80^\circ$,
 although the polarization degree is as high as $P\sim 15\%$ for 
 $\phi\lesssim 15^\circ$,
 the degree of polarization is suppressed to $P\lesssim 2\%$ for $\phi\gtrsim 75^\circ$.
This is because the direction of $(\theta,\phi)=(90^\circ,0^\circ)$
 is perpendicular to the global magnetic field,
 while that of $(\theta,\phi)=(90^\circ,90^\circ)$ is parallel to the global magnetic field.
This is consistent with the expectation for a uniform-density filament 
 threaded with a uniform magnetic field.
 
\subsection{Model with High Central Density}

Figures \ref{fig:rhoc300pol} and \ref{fig:rhoc300pol_all} show
 polarization patterns for Model B,
 which has the same parameter $R_0=2\,c_s/(4\pi G \rho_s)^{1/2}$
 and $\beta_0=1$ as that in the previous subsection,
 but with a different central density of $\rho_c=300\rho_s$. 
The upper panels of Figure \ref{fig:rhoc300pol} show the result for $\theta=30^\circ$.
Figure~\ref{fig:rhoc300pol}(a) with $(\theta,\phi)=(30^\circ,0^\circ)$ shows that the polarization 
 direction is perpendicular to the filament, which is similar to 
 the model with low central density (Figure~\ref{fig:rhoc10pol}(a)).
However, Figures \ref{fig:rhoc300pol}(b: $\phi=45^\circ$) and (c: $\phi=90^\circ$) reveal a
 clear difference from the corresponding models with low central density
 (Figures~\ref{fig:rhoc10pol} (b) and (c)).
Figure \ref{fig:rhoc10pol} has polarization vectors running from upper-left
 to lower-right (b) and parallel to the filament (c). 
However, Figures \ref{fig:rhoc300pol} (b) and (c) have 
 polarization vectors that are perpendicular to the filament,
 in a global sense.
By increasing $\phi$ from $\phi=0^\circ$ to $\phi=90^\circ$,
 $\alpha$ increases from $\alpha\sim 90^\circ$ to $\alpha\sim 180^\circ$ for
 the inner central region of the filament $|\xi| \lesssim 1\times 10^4 {\rm AU}$  
 (Figure~\ref{fig:rhoc300pol_all}(a)).
In contrast, the outer part ($|\xi| \gtrsim 1\times 10^4 {\rm AU}$)
 shows a different feature and $\alpha$ changes as
 $\alpha = 90^\circ$ ($\phi=0^\circ$),
 $\alpha \sim 50-90^\circ$ ($\phi=45^\circ$),
 and $\alpha \sim 90^\circ$ ($\phi=90^\circ$).
The outer part shows the polarization perpendicular to the filament ($\alpha\sim90^\circ$). 

Observation of the filament from the line-of-sight of $\theta=80^\circ$ reveals
 the polarization vector is also perpendicular to the filament
 (Figures~\ref{fig:rhoc300pol}(d)--(f)).
Figure \ref{fig:rhoc300pol_all}(d) shows that
 although the polarization direction angle $\alpha$ increases from
 $\alpha\simeq 90^\circ(\phi=0^\circ)$ 
 to $\alpha \simeq 180^\circ(\phi=90^\circ)$ 
 in the central part of the filament, $|\xi| \lesssim 5\times10^3 {\rm AU}$,
 $\alpha$ stays constant $\alpha\simeq 90^\circ$, irrespective of $\phi$ in the outer part of 
 $|\xi| \gtrsim 5\times10^3 {\rm AU}$.
Thus, Model B indicates a distinctly different polarization pattern from Model A,
 for both line-of-sights at $\theta=30^\circ$ and $\theta=80^\circ$.

The expected degree of polarization $P$ for Model B
 (Figures~\ref{fig:rhoc300pol_all}(c) and (f))
 is also very different from that of Model A
 (Figures~\ref{fig:rhoc10pol_all}(c) and (f)).
In Model A, $P$ decreases from 15\% ($\phi=0^\circ$) to 0\% ($\phi=90^\circ$),
 depending on $\phi$, in the case of $\theta=80^\circ$.
This is also observed in the central part of the filament,
 $|\xi|\lesssim 2\times10^4{\rm AU}$ in Model B. 
In contrast, for $|\xi|\gtrsim 2\times10^4{\rm AU}$, 
 $P\gtrsim 10\%$, irrespective of $\phi$ in Model B.
Therefore, even if the line-of-sight is parallel to the global magnetic field,
 the outer part of the filament for Model B indicates strong polarization,
in a direction perpendicular to the filament.

In summary, Model A and the inner part of Model B
 show similar polarization patterns.
However, the polarization pattern is different
 for the outer part of the filament for Model B.
The reason for this difference is clear.
Magnetic field lines threading the filament of Model A are straight.
In contrast, the magnetic field lines in the outer part of the filament of Model B
 are dragged inwardly near the equator,
 which induces a relatively strong $B_x$ component.
Considering the line-of-sight at $(\theta,\phi)=(80^\circ, 90^\circ)$,
 even when the filament is observed from the direction of the $y$-axis,
 the $B_x$ component, which is perpendicular to the line-of-sight,
 generates a certain amount of polarization.
   
\section{Discussion}
\subsection{Effect of Mass Loading}

\begin{table}
\begin{tabular}{lcccccc}
Model & $R_{0}$ & $\beta_0$ & $\rho_c$ & $\lambda_0$ & ${\cal N}$ & $D({\cal N})$\\
\hline
C1\ldots\ldots & $2\,c_s/(4\pi G \rho_s)^{1/2}$ & 0.1 & $19.2\rho_s$ & $3c_s^2/G$ & 0.1 & 1.03028\\
C2\ldots\ldots & $2\,c_s/(4\pi G \rho_s)^{1/2}$ & 0.1 & $30.54\rho_s$  & $3c_s^2/G$ & 1   & 1.27324\\
C3\ldots\ldots & $2\,c_s/(4\pi G \rho_s)^{1/2}$ & 0.1 & $416\rho_s$ & $3c_s^2/G$ & 6   & 2.1875\\
C4\ldots\ldots & $2\,c_s/(4\pi G \rho_s)^{1/2}$ & 0.1 & $416\rho_s$ & $3.76c_s^2/G$ & 1   & 1.27324\\
\hline
\end{tabular}
\caption{\label{tbl2:model-parameters}Model parameters for Figure~\ref{fig:R2b0.1N}.
Same as Table \ref{tbl1:model-parameters}, but where ${\cal N}$ and $D({\cal N})$
 represent the mass concentration index defined in Equation (\ref{eqn:generalizedN})
 and the degree of mass concentration defined in Equation (\ref{eqn:degreeofconcentration}), respectively.}
\end{table}

In this section, we compare the filaments with different mass loadings
 (mass distribution against magnetic flux).
In paper I, we assume a mass loading that is realized when a uniform-density
 cylinder with a density $\rho_0$ and a radius $R_0$
 is threaded by a uniform magnetic field $B_0$. 
In this model,
 the line-mass distribution $\lambda$, against the flux function $\Phi$, defined as
 the amount of magnetic flux counted
 from the central flux tube, is expressed as:
\be
\Difd{\lambda}{\Phi}= 2 \left(\rho_0\frac{R_0^2 }{\Phi_{\rm cl}}\right)
 \left[1-(\Phi/\Phi_{\rm cl})^2\right]^{1/2},
\ee
where $\Phi_{\rm cl}$ is the magnetic flux per unit length of a cloud,
 which is defined as 
\be
\Phi_{\rm cl}=R_0B_0,
\ee
 and the flux function
 $\Phi$ varies from $-\Phi_{\rm cl}$ to $+\Phi_{\rm cl}$.
This mass loading is extended to the following form:
\be
\Difd{\lambda}{\Phi}= 2 \left(\rho_0\frac{R_0^2 }{\Phi_{\rm cl}}\right)
 \left[1-(\Phi/\Phi_{\rm cl})^2\right]^{{\cal N}/2}.
\label{eqn:generalizedN}
\ee
Here, ${\cal N}$ represents the degree of mass concentration to the central
 magnetic flux tube and we call ${\cal N}$ here as the mass concentration index.
When ${\cal N}=0$,  $\Dif{\lambda}{\Phi}=$constant, irrespective of $\Phi$,
 which indicates a uniform mass loading:

\be
 \left. \Difd{\lambda}{\Phi}\right|_{\Phi=0}=\left. \Difd{\lambda}{\Phi}\right|_{\rm ave}\equiv\frac{\int_{-\Phi_{cl}}^{+\Phi_{cl}} \Difd{\lambda}{\Phi}d\Phi}{2\Phi_{cl}}.
\ee
By increasing the index ${\cal N}$,
 we are selecting the centrally concentrated mass loading 
 and the degree of mass concentration 
 \be
 D({\cal N})\equiv \left. \Difd{\lambda}{\Phi}\right|_{\Phi=0} / \left. \Difd{\lambda}{\Phi}\right|_{\rm ave}=\frac{2\Gamma[({\cal N}+3)/2]}{\pi^{1/2}\Gamma[({\cal N}+2)/2]},
\label{eqn:degreeofconcentration}
\ee
is an increasing function of ${\cal N}$, where $\Gamma$ represents the gamma function.
This ratio $D({\cal N})$ is tabulated in Table 1 of \citet{hanawa2015}.

Figure \ref{fig:R2b0.1N} shows three models (Models C1-C3) with different ${\cal N}$, 
 where the index ${\cal N}$ is chosen as
 ${\cal N}=0.1$ (Figure~\ref{fig:R2b0.1N}(a)), ${\cal N}=1$ (Figure~\ref{fig:R2b0.1N}(b)),
 and ${\cal N}=6$ (Figure~\ref{fig:R2b0.1N}(c)), respectively. However, the three models have the identical line-mass of $\lambda_0=3c_s^2/G$.
The parameters for these models are summarized
 in Table \ref{tbl2:model-parameters}.
By increasing ${\cal N}$, 
 the central density increases as $\rho_c=19.2\rho_s$ (Model C1),
 $\rho_c=30.54\rho_s$ (Model C2), and $\rho_c=416\rho_s$ (Model C3).
See also Figure 5 of \citet{hanawa2015}.
The gas is more concentrated toward the central magnetic flux tube; therefore,
 the gravity must be counter-balanced by the thermal pressure gradient,
 and thus the central density $\rho_c$ increases.
Figure \ref{fig:R2b0.1N} shows that
 the area of the cross-cut is also contracted when a larger ${\cal N}$ is selected.
The central concentration factor $D$,
 increases\footnote{$D=1$ for ${\cal N}=0$.}
 from $D=1.0303$ of ${\cal N}=0.1$
 to $D=2.1875$ of ${\cal N}=6$.

\begin{figure}
\noindent
\hspace*{2cm}(a)\hspace*{7cm}(b)\\
\epsscale{1}
%\begin{center}
%\plottwo{ps/Paper1_f5a.eps}{ps/Paper1_f5c.eps}
\plottwo{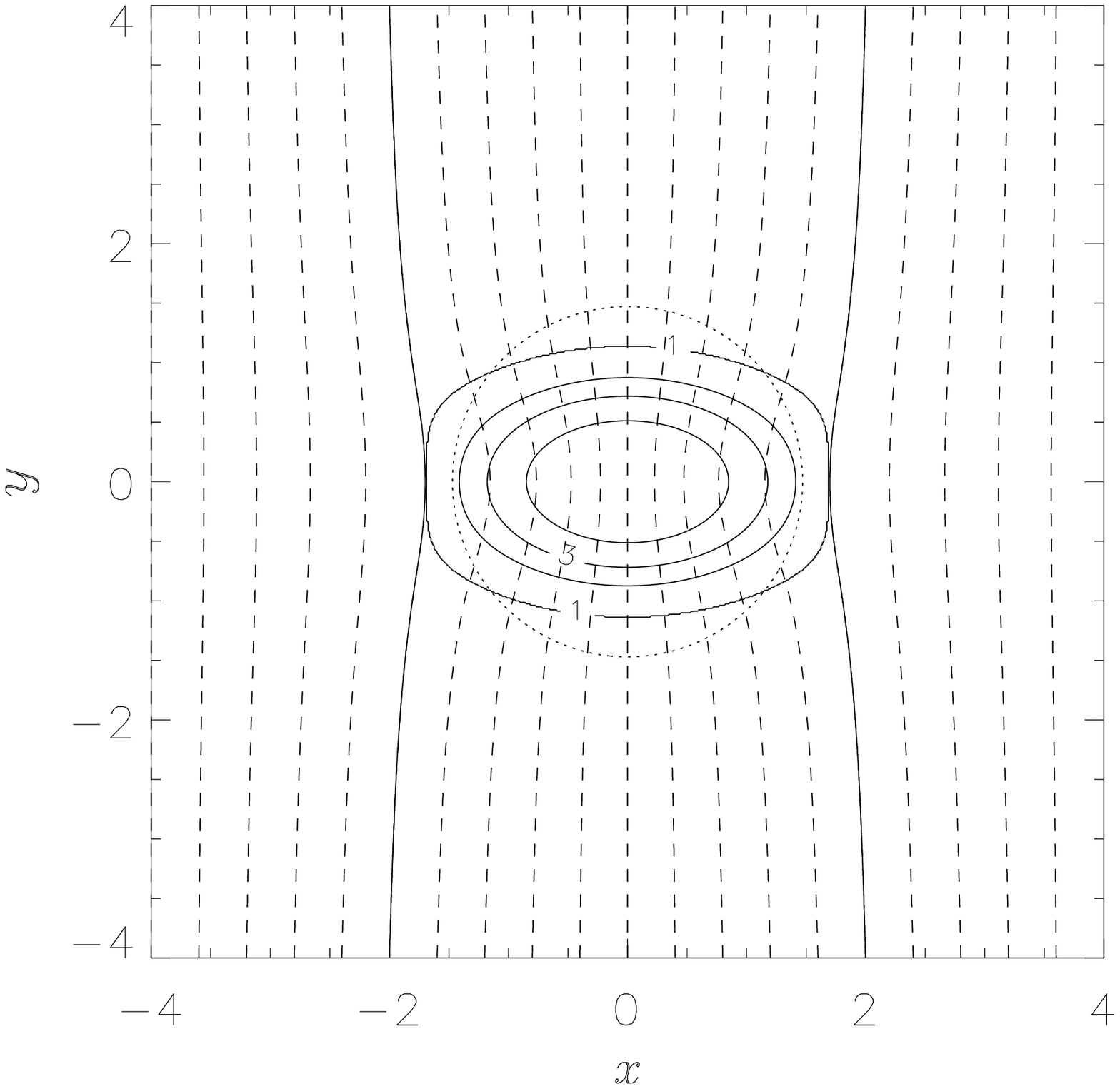}{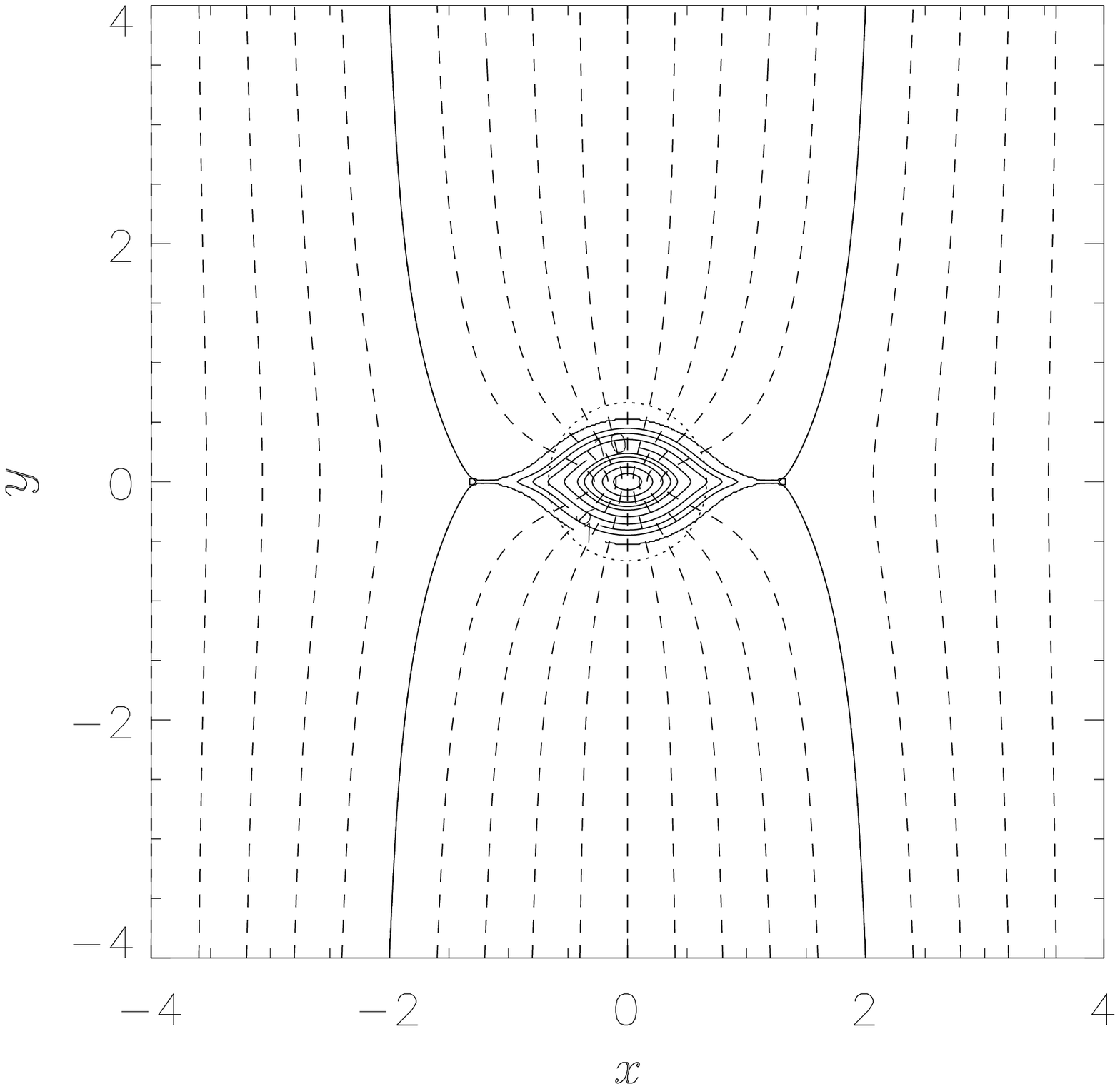}
%\end{center}
\caption{\label{fig:Paper1Fig5}
Magnetohydrostatic structures of models with $R_0=2\,c_s/(4\pi G \rho_s)^{1/2}$
 and $\beta_0=1$.
Two models are shown with different density contrasts between the
 center and the surface; (a) $\rho_c/\rho_s=10$ (Model A) and (b) $\rho_c/\rho_s=300$ (Model B),
which are taken from Figures 5(a) and (c) of paper I.
Closed solid lines represent the density contours,
 where the contour levels are selected as 1, 2, 3, 5, 10, 20, 30, 50, 100, and 200 
$\times\rho_s$.
Dashed lines running vertically represent the magnetic field lines,
 and the solid lines represent special magnetic field lines in contact with
 the cloud surface, $\rho=\rho_s$.
The dotted circle is shown to indicate the radius of the non-magnetized filament
 with the identical density contrast.
Model parameters shown in this figure are summarized in Table \ref{tbl1:model-parameters}.
The $x$- and $y$-axes represent the distance normalized with the scale-height
 $c_s/(4\pi G\rho_s)^{1/2}$.}
\end{figure}

\begin{figure}
\epsscale{1}
%\plotone{ps/R2b1N641rhoc10L.eps}
\plotone{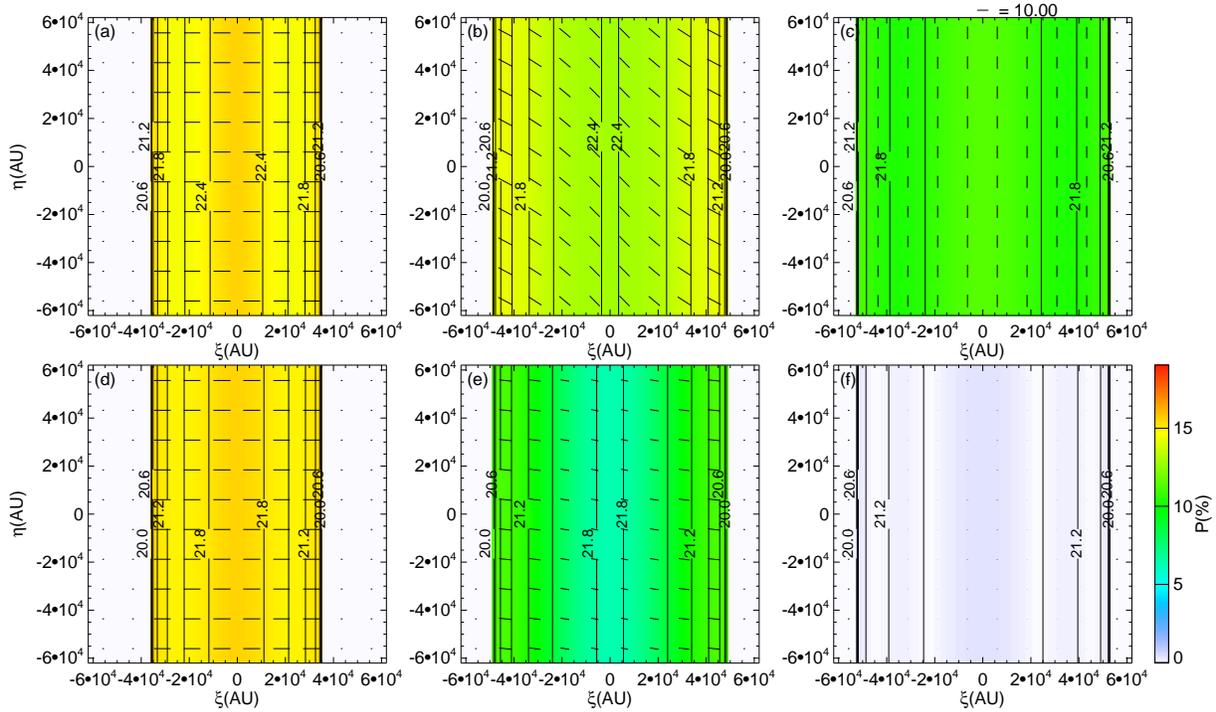}
%\plottwo{ps/R2b1N641rhoc10th30phi000.eps}{ps/R2b1N641rhoc10th80phi000.eps}\\
%\plottwo{ps/R2b1N641rhoc10th30phi045.eps}{ps/R2b1N641rhoc10th80phi045.eps}\\
%\plottwo{ps/R2b1N641rhoc10th30phi090.eps}{ps/R2b1N641rhoc10th80phi090.eps}\\
\caption{Expected polarization for the $R_0=2\,c_s/(4\pi G \rho_s)^{1/2}$, $\beta_0=1$ and $\rho_c=10\rho_s$ model (Model A).
Upper and lower panels correspond to models
 where the line-of-sight is selected with the angle from the filament axis
 at $\theta=30^\circ$ and $\theta=80^\circ$, respectively.
Left, center, and right panels represent the cases of $\phi=0^\circ$,
 $45^\circ$, and $90^\circ$, respectively.
Black bars represent the direction of the B-vector for the electromagnetic wave
 (polarization vector).
%False color and white contour lines represent the degree of polarization and
% black contour lines indicate the iso-column-density lines.
False color represents the degree of polarization and
 black contour lines indicate the iso-column-density lines 
 {\bf with a logarithmic step of $\Delta \log \Sigma=0.3$}.
\label{fig:rhoc10pol}}
\end{figure}

\begin{figure}
\epsscale{1}
%\plotone{ps/R2b1N641rhoc10th30+80L.eps}
\plotone{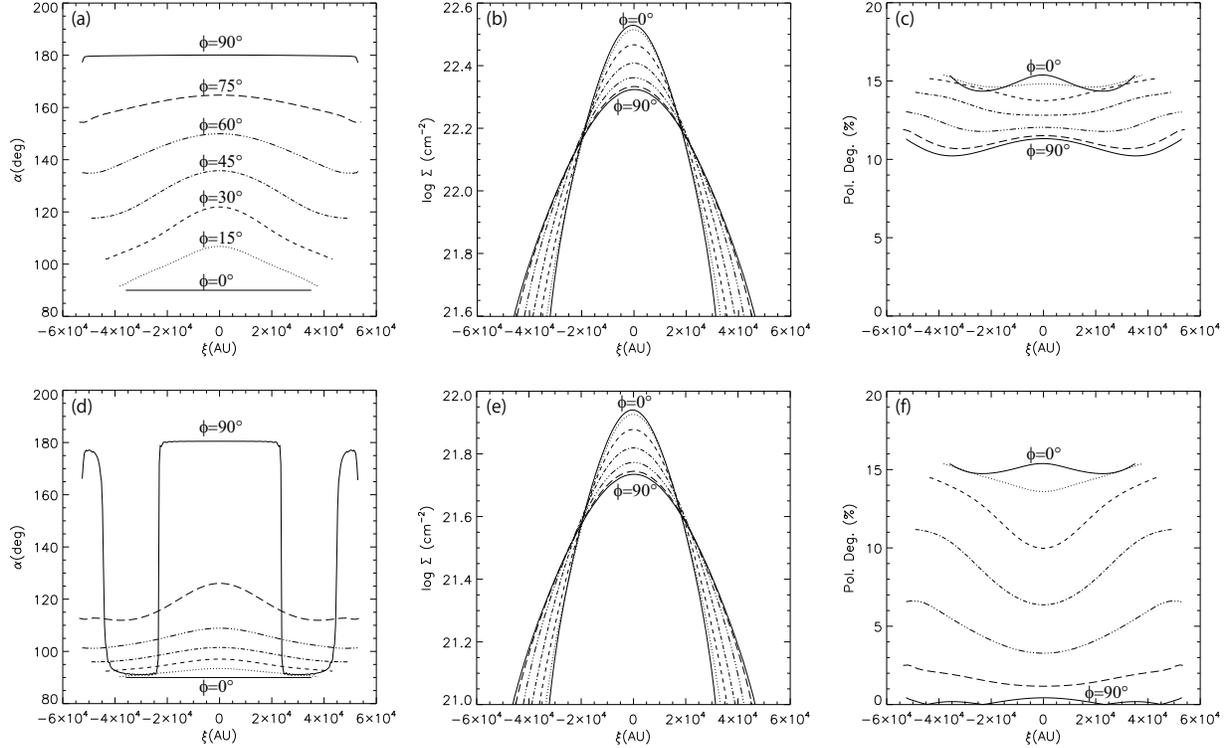}
%\plottwo{ps/R2b1N641rhoc10th30_PA.eps}{ps/R2b1N641rhoc10th80_PA.eps}\\
%\plottwo{ps/R2b1N641rhoc10th30_Sigma.eps}{ps/R2b1N641rhoc10th80_Sigma.eps}\\
%\plottwo{ps/R2b1N641rhoc10th30_PI.eps}{ps/R2b1N641rhoc10th80_PI.eps}\\
\caption{Expected polarization for the $R_0=2\,c_s/(4\pi G \rho_s)^{1/2}$, $\beta_0=1$,
 and $\rho_c=10\rho_s$ model (Model A).
Angle between the filament axis and polarization B-vector (left panels),
column density (center panels), and degree of polarization (right panels) are
plotted against the distance from the center of the filament.
Upper and lower panels correspond to the models
 where the line-of-sight is selected with the angle from the filament axis
 at $\theta=30^\circ$ and $\theta=80^\circ$, respectively.
Seven models with 
$\phi=0^\circ$ (solid line),
$\phi=15^\circ$ (dotted line),
$\phi=30^\circ$ (dashed line),
$\phi=45^\circ$ (dash-dotted line),
$\phi=60^\circ$ (two-dot chain line),
$\phi=75^\circ$ (long dashed line), and
$\phi=90^\circ$ (solid line) are shown.
\label{fig:rhoc10pol_all}}
\end{figure}

\begin{figure}
\epsscale{1.0}
%\plotone{ps/R2b1N641rhoc300L.eps}
\plotone{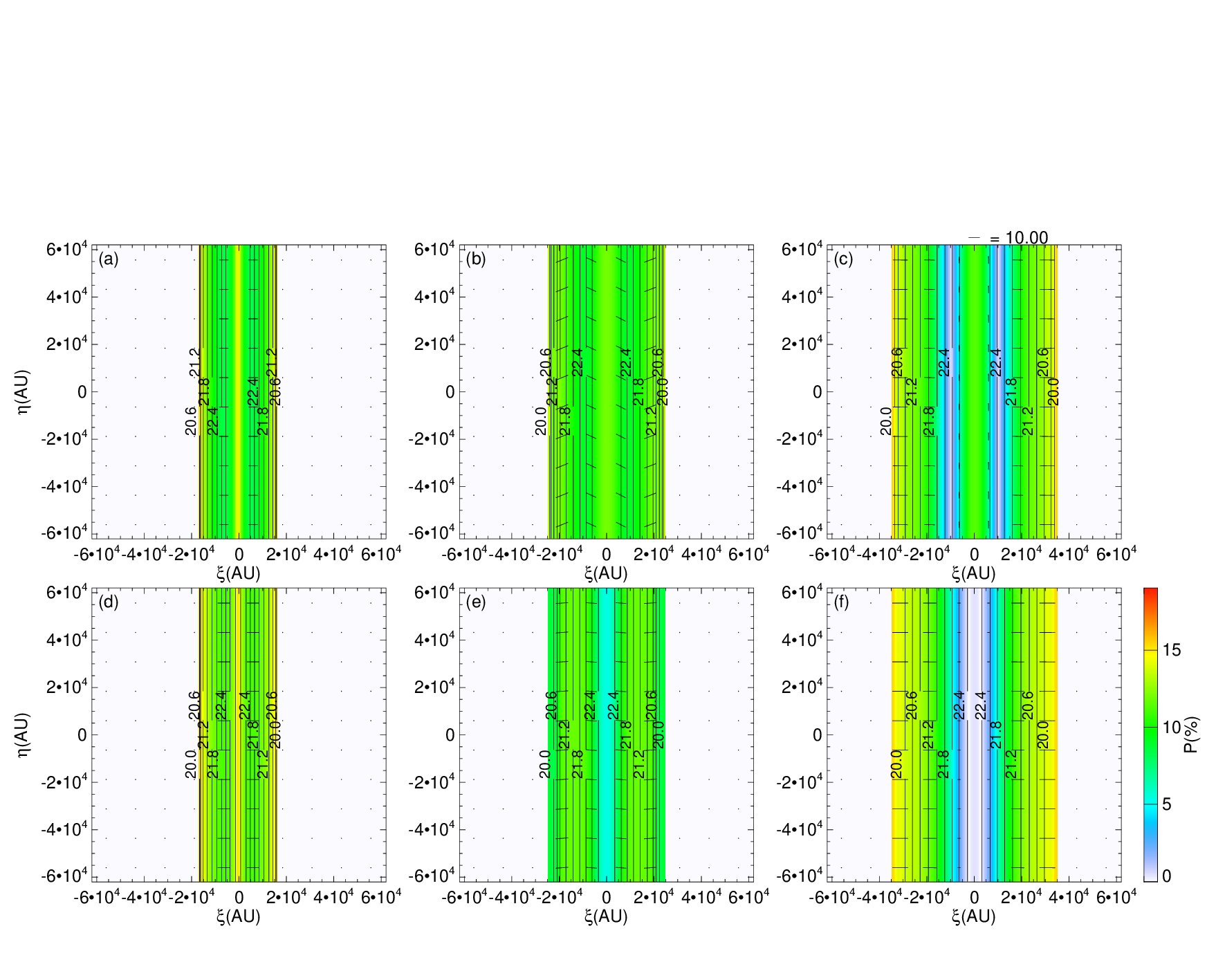}
%\plottwo{ps/R2b1N641rhoc300th30phi000.eps}{ps/R2b1N641rhoc300th80phi000.eps}\\
%\plottwo{ps/R2b1N641rhoc300th30phi045.eps}{ps/R2b1N641rhoc300th80phi045.eps}\\
%\plottwo{ps/R2b1N641rhoc300th30phi090.eps}{ps/R2b1N641rhoc300th80phi090.eps}\\
\caption{As for Figure~\ref{fig:rhoc10pol} but for the 
$R_0=2\,c_s/(4\pi G \rho_s)^{1/2}$, $\beta_0=1$, and $\rho_c=300\rho_s$ model
(Model B). 
%Upper and lower panels correspond respectively to the models
% whose line-of-sights are chosen with the angle from the filament axis
% of $\theta=30^\circ$ and $\theta=80^\circ$.
%Left, center, and right panels represent respectively $\phi=0^\circ$,
% $45^\circ$, and  $90^\circ$.
\label{fig:rhoc300pol}}
\end{figure}

\begin{figure}
\epsscale{1}
%\plotone{ps/R2b1N641rhoc300th30+80L.eps}
\plotone{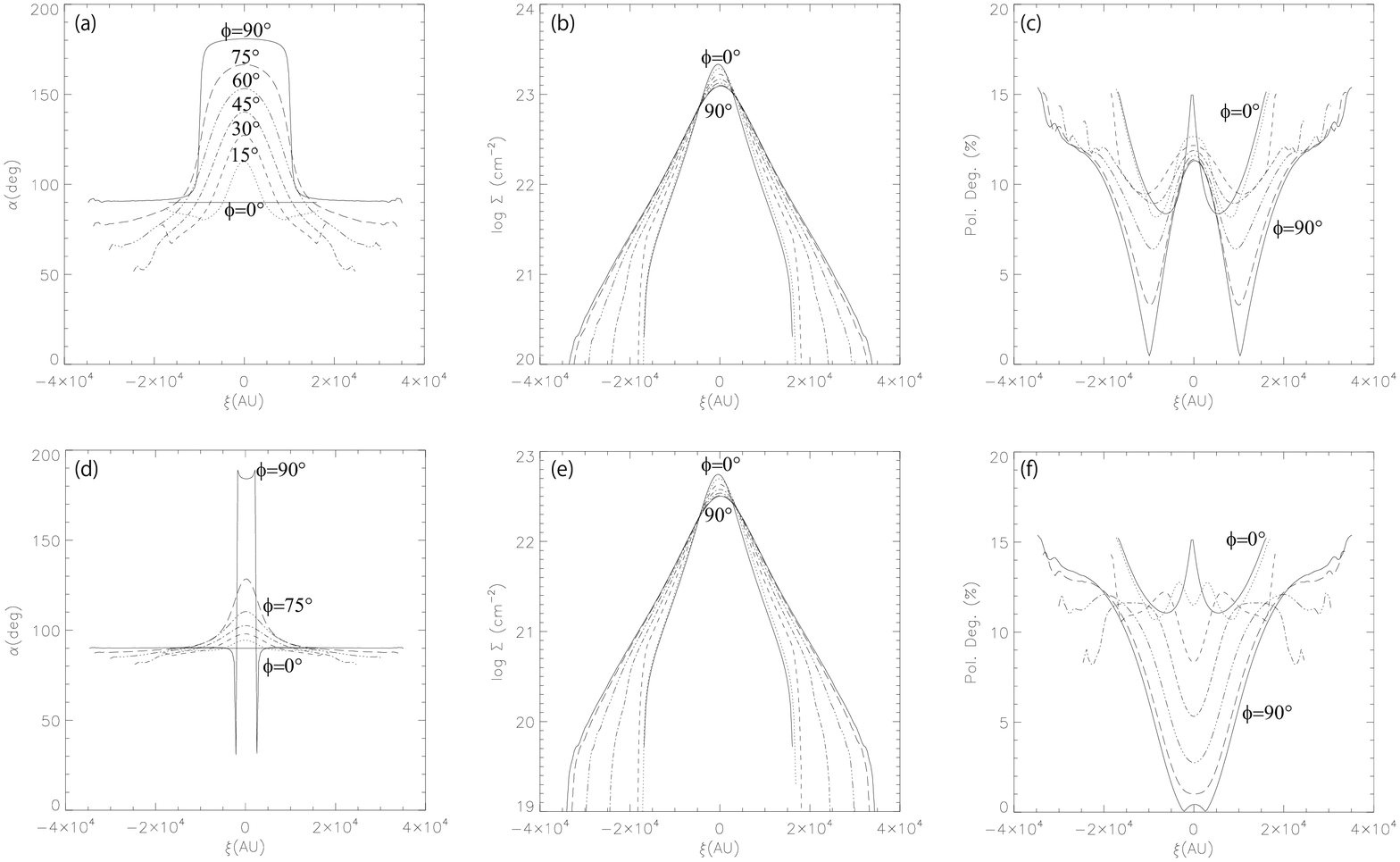}
%\plottwo{ps/R2b1N641rhoc300th30_PA.eps}{ps/R2b1N641rhoc300th80_PA.eps}\\
%\plottwo{ps/R2b1N641rhoc300th30_Sigma.eps}{ps/R2b1N641rhoc300th80_Sigma.eps}\\
%\plottwo{ps/R2b1N641rhoc300th30_PI.eps}{ps/R2b1N641rhoc300th80_PI.eps}\\
\caption{As for Figure~\ref{fig:rhoc10pol_all} but for the 
$R_0=2\,c_s/(4\pi G \rho_s)^{1/2}$, $\beta_0=1$, and $\rho_c=300\rho_s$ model
(Model B).
%Angle between the filament axis and polarization B-vector (left),
%column density (center), and polarized intensity (right) are
%plotted against the distance from the center of the filament.
%Upper and lower panels correspond respectively to the models
% whose line-of-sights are chosen with the angle from the filament axis
% of $\theta=30$ and $\theta=80$.
%Seven models with 
%$\phi=0^\circ$ (solid line),
%$\phi=15^\circ$ (dotted line),
%$\phi=30^\circ$ (dashed line),
%$\phi=45^\circ$ (dash-dotted line),
%$\phi=60^\circ$ (two-dot chain line),
%$\phi=75^\circ$ (long dashed line), and
%$\phi=90^\circ$ (solid line) are shown.
\label{fig:rhoc300pol_all}}
\end{figure}

\begin{figure}
\noindent
\hspace*{2cm}(a)\hspace*{7cm}(b)\\
\epsscale{1}
%\plottwo{ps/N0.1R2b0.1rhoc=5.87.eps}{ps/N1R2b0.1rhoc=8.01.eps}\\
%\plottwo{ps/N0.1R2b0.1rhoc=19.2.eps}{ps/N1R2b0.1rhoc=30.5.eps}\\
\plottwo{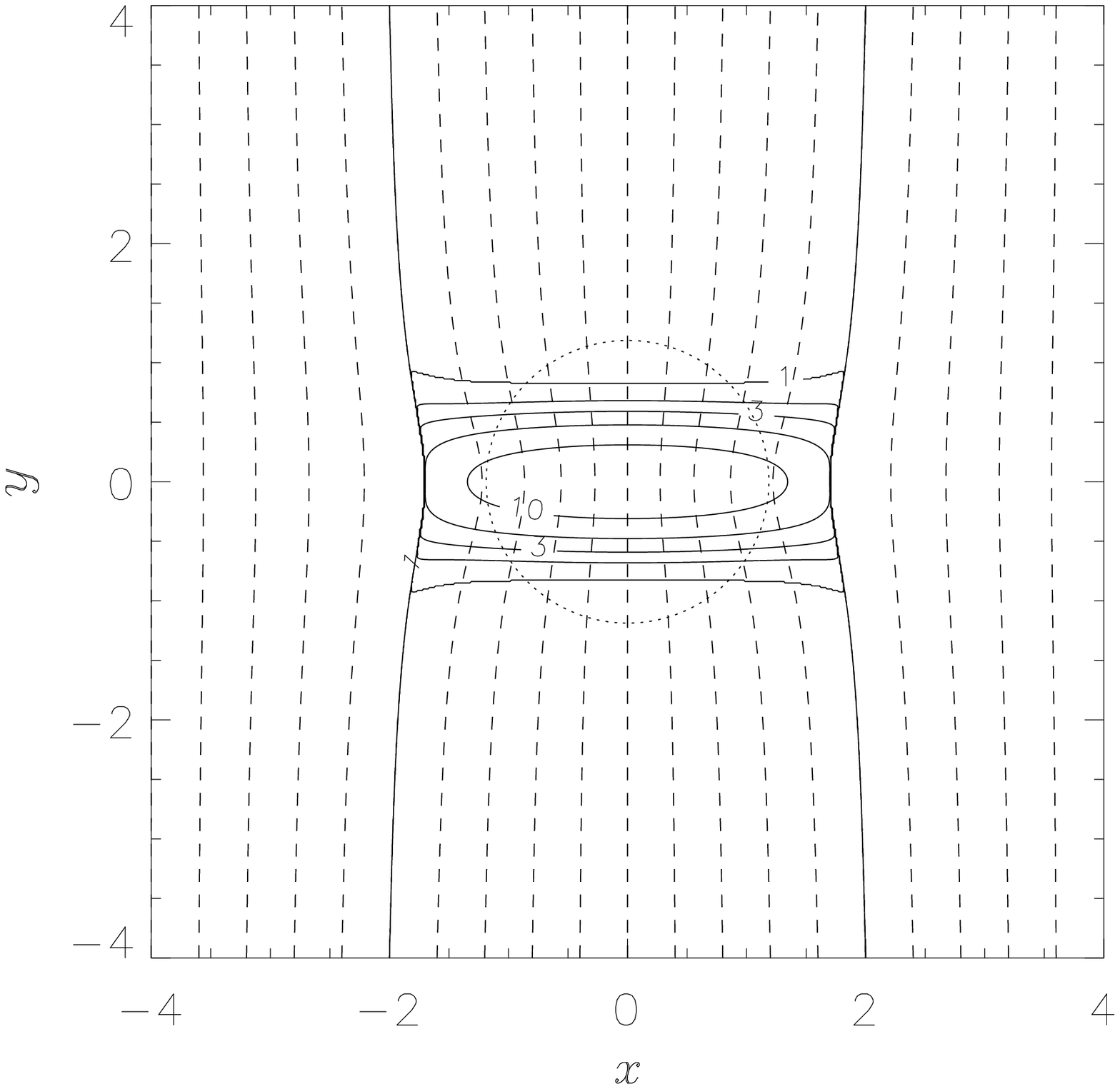}{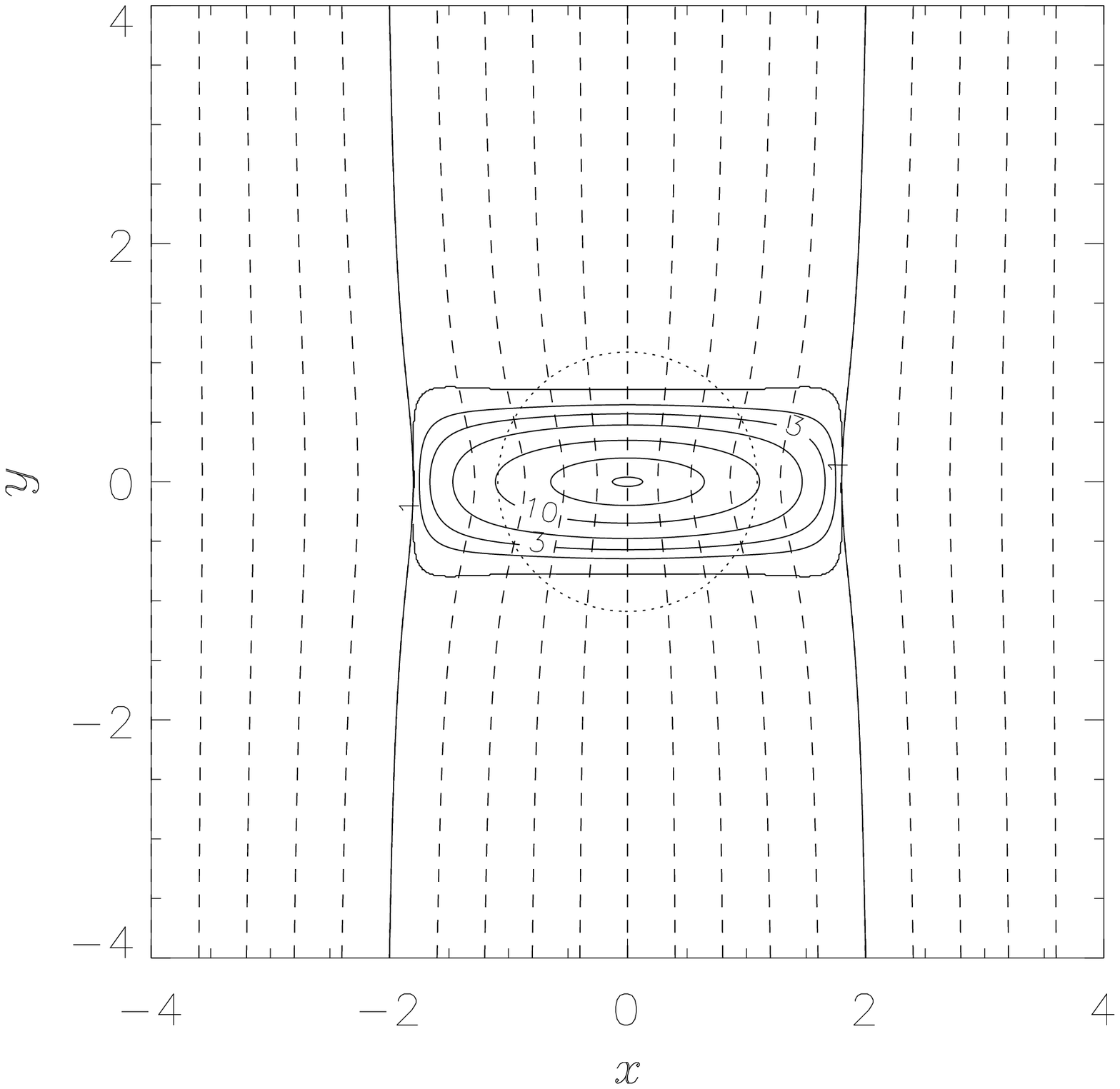}\\
\hspace*{2cm}(c)\hspace*{7cm}(d)\\
%\hspace*{4cm}(c)\\
 \epsscale{1}
%\plotone{ps/N10R2b0.1rhoc=29.3.eps}
%\plottwo{ps/N6R2b0.1rhoc=416..eps}{ps/N1R2b0.1rhoc=416..eps}
\plottwo{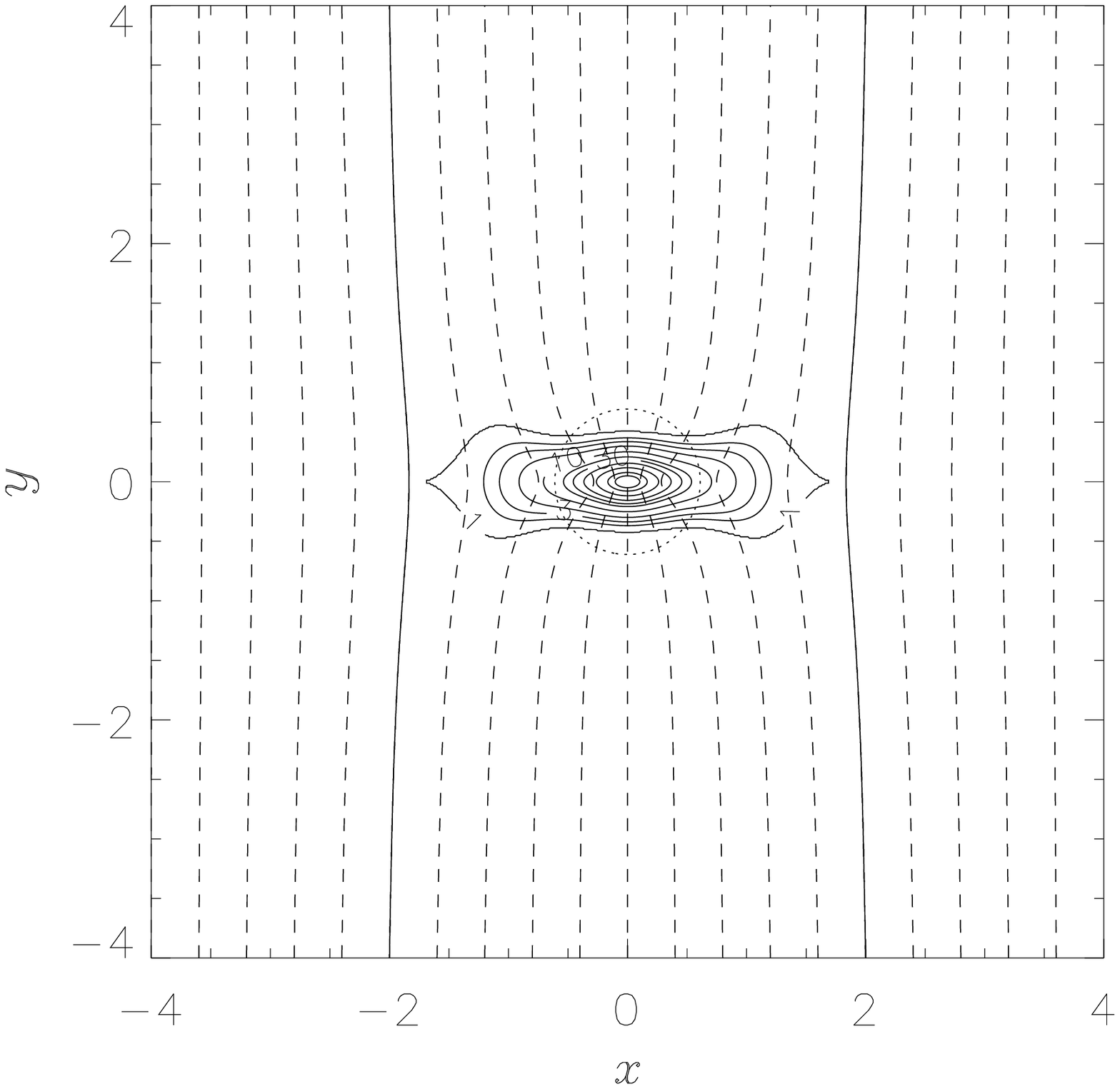}{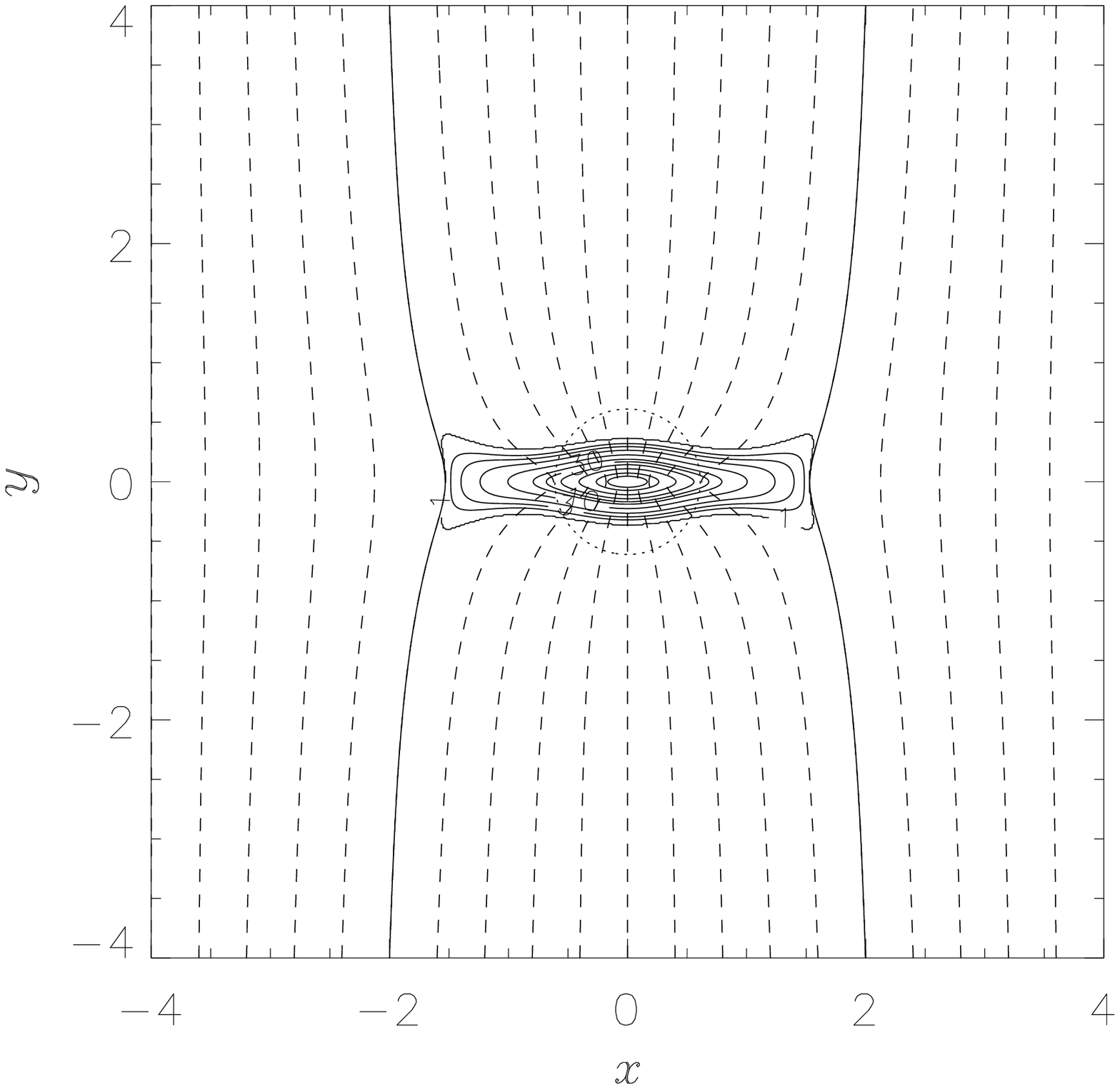}
\caption{Structure of hydrostatic filaments with the same line-mass
but different mass-loadings.
%Three filaments have the identical line-mass of $\lambda=2c_s^2/G$. 
Models C1 (a), C2 (b), and C3 (c) have identical line-mass of $\lambda_0=3c_s^2/G$. 
%Parameters of each panel are (a) ${\cal N}=0.1$ and $\rho_c=5.87\rho_s$,
%(b) ${\cal N}=1$ and $\rho_c=8.01\rho_s$, and
%(c) ${\cal N}=10$ and $\rho_c=29.3\rho_s$.
Parameters of each panel are ({\it a}) ${\cal N}=0.1$ and $\rho_c=19.2\rho_s$,
({\it b}) ${\cal N}=1$ and $\rho_c=30.54\rho_s$, and
({\it c}) ${\cal N}=6$ and $\rho_c=416\rho_s$.
Model C4 in ({\it d}) has the same central density of $\rho_c=416\rho_s$
 as Model C3 ({\it c}) but different ${\cal N}=1$, and thus line-mass $\lambda_0=3.76c_s^2/G$. 
Solid and dashed lines represent 
 the density contours and magnetic field lines, respectively, as in
 Figure~\ref{fig:Paper1Fig5}.
Centrally concentrated mass-loading (increasing ${\cal N}$ 
 from (a) to (c)) induces higher central density.
The parameters for these models are given in Table \ref{tbl2:model-parameters}.
\label{fig:R2b0.1N}}
\end{figure}

In $\S$ \ref{sec:result}, Figure \ref{fig:Paper1Fig5} shows that
 Model B with high central density (Figure~\ref{fig:Paper1Fig5}(b)) has magnetic field lines
 that are heavily squeezed toward the center near the equator ($y=0$),
 compared with Model A with a low central density (Figure~\ref{fig:Paper1Fig5}(a)).
However, Model C3 with a high central density shown in Figure \ref{fig:R2b0.1N}(c)
 has a magnetic field structure similar to Models C1 and C2
 with lower central densities
 (Figures~\ref{fig:R2b0.1N}(a) and (b)),
 especially for the outer part of the filament
 ($|x|\gtrsim 1$ in nondimensional distance).
This is clearly shown by a comparison of Figures \ref{fig:R2b0.1N}(c) and (d),
 both of which have the same central density $\rho_c=416\rho_s$ but
a different mass-loading index ${\cal N}$ and line-mass $\lambda_0$
 (for Model C3 of Figure~\ref{fig:R2b0.1N}(c), ${\cal N}=6$ and $\lambda_0=3c_s^2/G$ were selected,
 while for Model C4 in Figure~\ref{fig:R2b0.1N}(d), ${\cal N}=1$ and $\lambda_0=3.76c_s^2/G$ were selected). 
Although the magnetic field lines are dragged inwardly in both models,
 the field lines in Model C4 are squeezed toward the center
 more strongly than those of Model C3.
Model C3 has a more centrally concentrated mass loading
 and 
 the magnetic field is stored in the outer part of the filament.
Thus, the magnetic field lines run relatively straight in this model.
In conclusion, it is shown that
 the pattern of magnetic field lines is affected by how the mass is distributed 
 against the magnetic flux (mass loading)
 and by the center-to-surface density ratio (or the line-mass $\lambda_0$).
By increasing the mass concentration index ${\cal N}$,
 the magnetic field lines run more straight.

\subsection{Does the Polarization Pattern Depend on Mass Loading?}
\begin{figure}
\epsscale{1}
%\plotone{ps/N6R2b0.1rho416L.eps}
\plotone{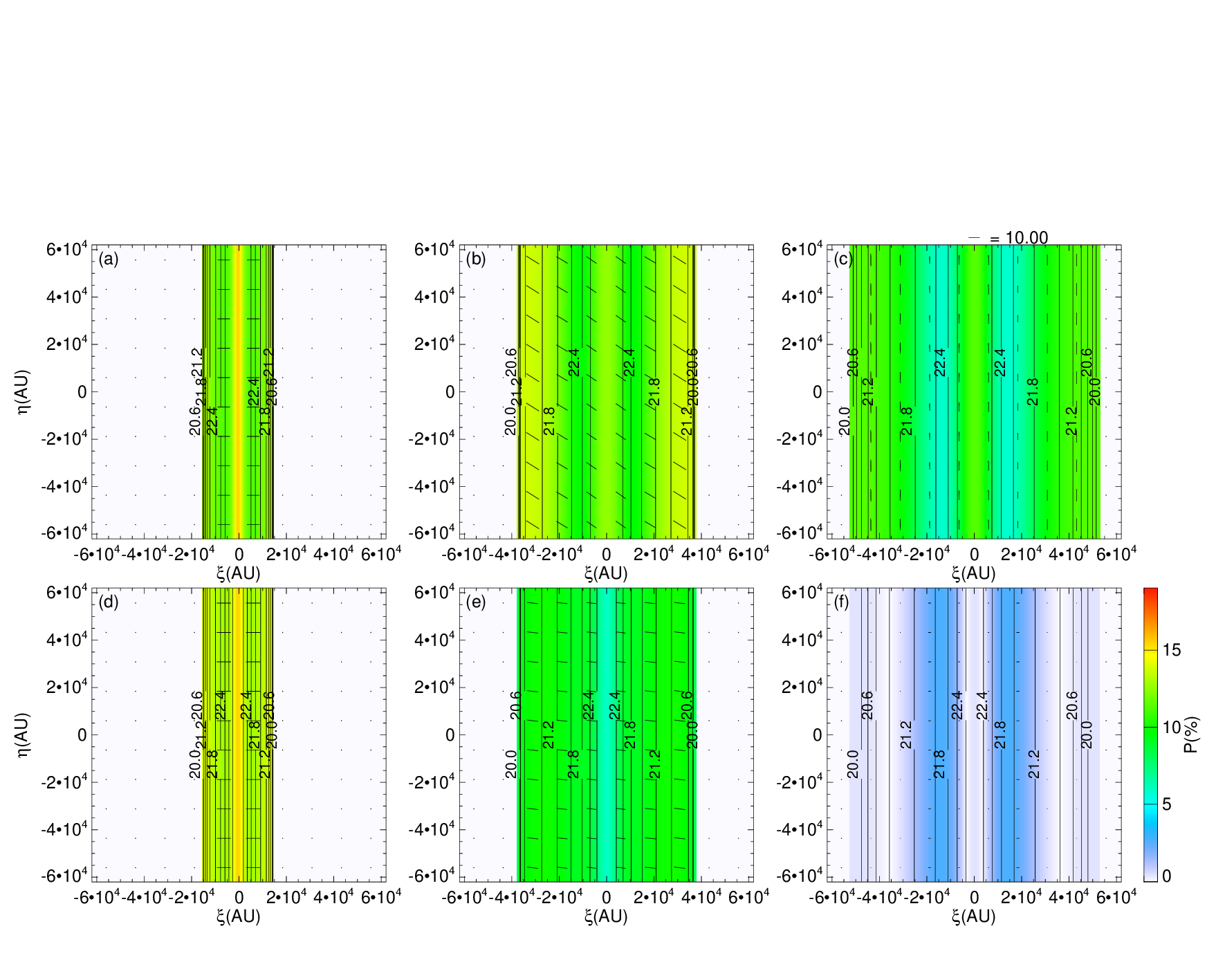}
\caption{As for Figure~\ref{fig:rhoc10pol} but for the 
$R_0=2\,c_s/(4\pi G \rho_s)^{1/2}$, $\beta_0=0.1$, ${\cal N}=6$,
 and $\rho_c=416\rho_s$ model (Model C3).
%Upper and lower panels correspond respectively to the models
% whose line-of-sights are chosen with the angle from the filament axis
% of $\theta=30^\circ$ and $\theta=80^\circ$.
%Left, center, and right panels represent respectively $\phi=0^\circ$,
% $45^\circ$, and  $90^\circ$.
\label{fig:N6pol}}
\end{figure}

As shown in $\S$4, 
the configuration of magnetic field lines is affected
 not only by the center-to-surface density ratio $\rho_c/\rho_s$ (paper I),
 but also by the mass-loading (or the mass-concentration index, ${\cal N}$).
The expected polarization pattern is also affected by the mass-loading.
Figure \ref{fig:N6pol} shows the expected polarization pattern for 
 Model C3 of Figure~\ref{fig:R2b0.1N}(c), which has a relatively large central density
 $\rho_c=416\rho_s$, but a large mass concentration index ${\cal N}=6$. 
Observation of the filament from a line-of-sight in the $x-z$ plane 
 (perpendicular to the global magnetic field), such as 
 $(\theta,\phi)=(30^\circ,0^\circ)$ (Figure~\ref{fig:N6pol}(a))
 and $(80^\circ,0^\circ)$ (Figure~\ref{fig:N6pol}(d)),
 we observe the polarization vector to be perpendicular to the filament,
 which is similar to Figures \ref{fig:rhoc10pol}(a) and (d) 
 and Figures \ref{fig:rhoc300pol}(a) and (d).
The same polarization pattern is observed in the case of
 $(\theta,\phi)=(80^\circ,45^\circ)$ in all 
 Figures \ref{fig:rhoc10pol}, \ref{fig:rhoc300pol}, and \ref{fig:N6pol}.
 
For lines-of-sight with $(\theta,\phi)=(30^\circ,45^\circ)$ (Figure~\ref{fig:N6pol}(b)),
 the polarization vector is running from the upper-left to the lower-right,
 which is similar to Figure~\ref{fig:rhoc10pol}(b) but different  
 from Figure~\ref{fig:rhoc300pol}(b).
Figure \ref{fig:N6pol}(c), in which the polarization vector is 
 parallel to the filament, does not resemble Figure \ref{fig:rhoc300pol}(c)
 but does resemble Figure \ref{fig:rhoc10pol}(c).
Observation of the filament from almost the direction of global magnetic field,
 $(\theta,\phi)=(80^\circ,90^\circ)$ (f), reveals a low degree of polarization.
This is not observed in Figure \ref{fig:rhoc300pol}(f),
 but is evident in Figure \ref{fig:rhoc10pol}(f).
In summary, Model C3 has a polarization pattern similar to Model A 
 (${\cal N}=1$ and low central density model), but
 not similar to Model B (${\cal N}=1$ and high central density model).
This clearly shows that
 the models with a large mass concentration index ${\cal N}$ have
 straight magnetic field lines,
 even near the equator of the outer part.
This gives a polarization pattern similar to
 Model A, but not similar to Model B,
 which indicates that the observed polarization pattern is
 affected not only by the center-to-surface density ratio $\rho_c/\rho_s$,
 but also by mass concentration index, ${\cal N}$.
Even if the central density is high, as in Model C3, the magnetic field lines
 are relatively straight, which induces the polarization pattern
 expected for a filament threaded by a straight magnetic field.

\subsection{Distribution of the Angles between Polarization Vector and the Filament Axis}

\begin{figure}
\epsscale{0.60}
\plotone{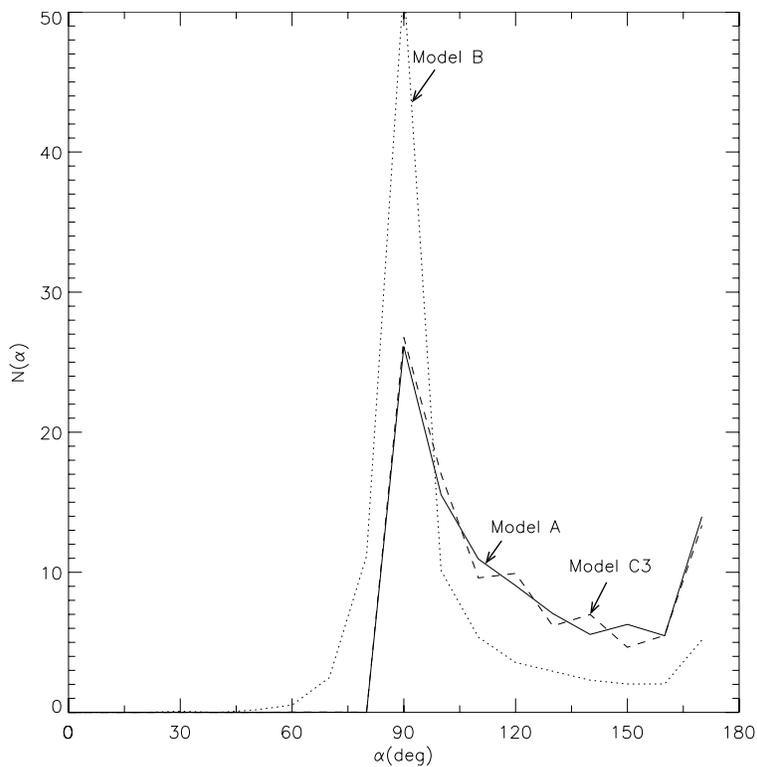}
\caption{\label{fig:N(alpha)}Distribution of the angle between polarization B-vectors and the filament axis, $\alpha$.
Solid, dotted, and dashed lines represent, respectively, Models A, B, and C3.
The $x$- and $y$-axes indicate $\alpha$ (deg) and the angle distribution $N(\alpha)$ in
 arbitrary unit.
This shows that the angle $\alpha$ is concentrated to $\alpha\simeq 90^\circ$ in Model B,
 which has a high central density.
Also in Models A and C3, the angle $\alpha$ is concentrated around $\alpha\simeq 90^\circ$.
However, the distributions are more uniform compared with Model B, 
 and have second peaks around $\simeq 180^\circ$.     
}
\end{figure}

The Planck polarization observation has indicated the distribution of angles between
 the polarization B-vector and the direction of iso-column density contours, $\varphi$ \citep{planck2015}.
Since the angle $\varphi$ corresponds to $\alpha$ of this paper,
 we calculate the distribution of angle $\alpha$.  
For each model, the number of grids whose angles equal to $\alpha$ is calculated.
Since this number of grids also depends on the direction of line-of-sight or $\theta$ and $\phi$,
 we express this as $n(\alpha; \theta, \phi)$.
If we assume the line-of-sight direction is randomly chosen,
 the expected distribution of $\alpha$ is obtained as follows: 
\begin{equation}
 N(\alpha)=\frac{\int_{\theta=0^\circ}^{\theta=90^\circ}\int_{\phi=0^\circ}^{\phi=180^\circ} n(\alpha; \theta, \phi)\sin\theta\, d\theta\, d\phi}{\int_{\theta=0^\circ}^{\theta=90^\circ}\int_{\phi=0^\circ}^{\phi=180^\circ} \sin\theta\, d\theta\, d\phi}.
\end{equation}   
In Figure \ref{fig:N(alpha)}, we plot $N(\alpha)$ for three models,
 Models A (Figure \ref{fig:rhoc10pol}), B (Figure \ref{fig:rhoc300pol}),
 and C3 (Figure \ref{fig:N6pol}).
To obtain the expected distribution of angle $n(\alpha; \theta, \phi)$,
 we do not take into account of the polarization degree $P$ and the polarized intensity.
However we only count the grids where the column density exceeds $10^{21} {\rm H_2\,cm^{-2}}$. 

Figure \ref{fig:N(alpha)} shows that
 all the three models have distribution function whose peaks are located
 around $\alpha\simeq 90^\circ$,
 which is consistent with the polarization observation with Planck
 seen in such as Taurus, Lupus, and Chamaeleon-Musca. 
Model B has a strongly concentrated distribution around $\alpha\simeq 90^\circ$,
 while Models A and C3 have more uniform distributions and have another peak
 around $\alpha\simeq 180^\circ$.

Figure \ref{fig:N(alpha)} shows that
 even if the magnetic field is running perpendicular to the filament in three dimension,
 some clouds may be observed with $\alpha\simeq 0^\circ$
 (filaments are aligned to the magnetic field).
Thus, we should pay attention to the projection effect in reconstructing the three dimensional
 configuration of the filament.      

\section{Summary}
We have identified two types of polarization patterns
 from mock observation of magnetohydrostatic filaments perpendicularly threaded by magnetic field.
When the center-to-surface density ratio for the filament is small,
 a pattern is realized
 in which the B-vector is perpendicular to the filament, when the filament 
 is observed from the line-of-sight perpendicular to the magnetic field.
However, when the filament is observed from the direction of the magnetic field,
 the observed degree of polarization is expected to be very low.
This pattern is similar to that expected for a filament with uniform density
 and uniform magnetic field.
This is also expected for a filament with high central density,
 if the mass concentration index ${\cal N}$ is large (gas mass is concentrated
 toward the central magnetic flux tube), because the magnetic field lines are
 globally straight also in this case.

In contrast, another pattern is expected for a filament with both a high
 center-to-surface density ratio and a low mass concentration
 index ${\cal N}\sim 1$.
In this pattern, the B-vector is observed perpendicular to the filament,
 even when the filament is observed from the direction of the magnetic field.
This may explain why filaments are often associated with perpendicular
 magnetic field lines. 
   
\section*{}
This work was supported in part by the Grant-in-Aid
for Scientific Research (A) (No. 21244021) from the Japan Society for the Promotion of Science (JSPS), and
by HPCI Strategic Program of the Japanese Ministry of Education, Culture, Sports, Science and Technology (MEXT). Numerical
computations were conducted, in part, on Cray XT4 and Cray
XC30 computers at the Center for Computational Astrophysics (CfCA) at
the National Astronomical Observatory of Japan.

\clearpage

\clearpage
%%%%%%%%%%%%%%%%%%%%%%%%%%%%%%%%%%%%%%%%%%%%%%%%%%%%%%%%%%%%%%%%%%%%%%%%%%
%  FIG.1
%%%%%%%%%%%%%%%%%%%%%%%%%%%%%%%%%%%%%%%%%%%%%%%%%%%%%%%%%%%%%%%%%%%%%%%%%%

%%%%%%%%%%%%%%%%%%%%%%%%%%%%%%%%%%%%%%%%%%%%%%%%%%%%%%%%%%%%%%%%%%%%%%%%%%
%  FIG.2
%%%%%%%%%%%%%%%%%%%%%%%%%%%%%%%%%%%%%%%%%%%%%%%%%%%%%%%%%%%%%%%%%%%%%%%%%%

%%%%%%%%%%%%%%%%%%%%%%%%%%%%%%%%%%%%%%%%%%%%%%%%%%%%%%%%%%%%%%%%%%%%%%%%%%
%  FIG.3
%%%%%%%%%%%%%%%%%%%%%%%%%%%%%%%%%%%%%%%%%%%%%%%%%%%%%%%%%%%%%%%%%%%%%%%%%%

%%%%%%%%%%%%%%%%%%%%%%%%%%%%%%%%%%%%%%%%%%%%%%%%%%%%%%%%%%%%%%%%%%%%%%%%%%
%  FIG.4
%%%%%%%%%%%%%%%%%%%%%%%%%%%%%%%%%%%%%%%%%%%%%%%%%%%%%%%%%%%%%%%%%%%%%%%%%%

%%%%%%%%%%%%%%%%%%%%%%%%%%%%%%%%%%%%%%%%%%%%%%%%%%%%%%%%%%%%%%%%%%%%%%%%%%
%  FIG.5
%%%%%%%%%%%%%%%%%%%%%%%%%%%%%%%%%%%%%%%%%%%%%%%%%%%%%%%%%%%%%%%%%%%%%%%%%%

%%%%%%%%%%%%%%%%%%%%%%%%%%%%%%%%%%%%%%%%%%%%%%%%%%%%%%%%%%%%%%%%%%%%%%%%%%
%  FIG.6
%%%%%%%%%%%%%%%%%%%%%%%%%%%%%%%%%%%%%%%%%%%%%%%%%%%%%%%%%%%%%%%%%%%%%%%%%%

%%%%%%%%%%%%%%%%%%%%%%%%%%%%%%%%%%%%%%%%%%%%%%%%%%%%%%%%%%%%%%%%%%%%%%%%%%
%  FIG.7
%%%%%%%%%%%%%%%%%%%%%%%%%%%%%%%%%%%%%%%%%%%%%%%%%%%%%%%%%%%%%%%%%%%%%%%%%%

%%%%%%%%%%%%%%%%%%%%%%%%%%%%%%%%%%%%%%%%%%%%%%%%%%%%%%%%%%%%%%%%%%%%%%%%%%
%  FIG.8
%%%%%%%%%%%%%%%%%%%%%%%%%%%%%%%%%%%%%%%%%%%%%%%%%%%%%%%%%%%%%%%%%%%%%%%%%%

%%%%%%%%%%%%%%%%%%%%%%%%%%%%%%%%%%%%%%%%%%%%%%%%%%%%%%%%%%%%%%%%%%%%%%%%%%
%  FIG.8
%%%%%%%%%%%%%%%%%%%%%%%%%%%%%%%%%%%%%%%%%%%%%%%%%%%%%%%%%%%%%%%%%%%%%%%%%%
\end{document}